# Tribological Properties of Ultrananocrystalline Diamond Nanowire Thin Film: Influence of Sliding Ball Counterbodies

Revati Rani[1], N. Kumar[1*] and I-Nan Lin[2]

[1]Materials Science Group, Indira Gandhi Centre for Atomic Research, HBNI Kalpakkam 603102, INDIA

[2]Department of Physics, Tamkamg University, Tamsui 25137, TAIWAN

*Abstract-* Ultrananocrystalline Diamond Nanowire (UNCD NW) thin film was deposited on mirror polished silicon substrate (100) using Microwave Plasma Enhanced Chemical Vapor Deposition (MPECVD) System with optimized deposition parameters in $CH_4$ (6%)/$N_2$ plasma media. The film exhibited wire like morphology with randomly oriented and homogeneously distributed ultranano diamond grains separated by an interphase boundary of graphitic and amorphous carbon (a-C) phases. Micro-tribological studies of film were carried out against $Al_2O_3$, SiC and steel balls in ambient atmospheric conditions. Initially, the friction coefficient was found to be high for UNCD NW/SiC and UNCD NW/Steel sliding pairs which gradually decreased to low value. While, in UNCD NW/$Al_2O_3$ sliding combination, the ultralow value of friction coefficient was maintained throughout the whole sliding process. High wear resistant properties of the film were observed in UNCD NW/SiC and UNCD NW/Steel pairs. In UNCD NW/$Al_2O_3$ case, ball counterbody showed negligible wear dimension. Such kind of tribological behavior was attributed to the different type of mechanical and chemical interactions of ball counterbodies with UNCD NW thin film.

## I. INTRODUCTION

Low friction and high wear resistance of thin films are demanding for sustainable and energy efficient applications. Carbon based materials are beneficial in this regard due to their tunable microstructure and morphology and therefore, it is easy to manipulate film surface to make it tribologically compatible. Nanocrystalline/ultrananocrystalline diamond (NCD/UNCD) thin films show nearly vanishing friction and high wear resistant properties [1-3].

There are several factors which may influence the tribological response of thin films. One among them is sliding counterbodies [3]. In real tribology applications, thin films can undergo alternate sliding against metallic and ceramic counterpart. In this condition, the wear and chemical structure of sliding interface becomes complex which may affect overall tribological efficiency.

Therefore, it is important to evaluate the tribological performance of this film under various counterbody interactions.

In this study, metallic and ceramic balls were used to quantify the tribological properties of UNCD NW film.

## II. EXPERIMENTAL SECTION

### A. Film Deposition and Characterization

UNCD NW film was grown on mirror finished silicon (100) substrate using MPECVD system (2.45 GHz 6'' IPLAS-CYRANNUS). Before the deposition process, substrate was ultrasonicated in a methanol solution containing a mixture of nanodiamond powder (~5 nm) and titanium powder (325 nm) for 45 minutes to facilitate the generation of surface sites for nucleation process. The UNCD NW film was deposited using $CH_4$ (6%)/$N_2$ plasma with a microwave power of 1200 W, a pressure of 50 Torr and a substrate temperature of 700°C. Detailed deposition process is reported elsewhere [1].

The surface topography and roughness of film were analyzed using an atomic force microscope (AFM, Park XE-100). The local chemical structure of film surface was examined using Micro-Raman spectrometer (Andor SR-500i-C-R, wavelength 532 nm). The morphology of film was obtained using field emission scanning electron microscope (FE-SEM, Zeiss Supra 55). The hardness and elastic modulus of film were evaluated by nanoindentation (Triboindenter TI 950, USA) coupled with Berkovich diamond indenter with a tip curvature of 150 nm. A maximum load of 6 mN and a loading-unloading rate of 1.5 mN/minute were used. To avoid any substrate effect, the maximum penetration depth of indentation was kept well below the 1/10[th] of the film thickness. Rutherford Back Scattering technique with α particle as projectile (energy 3800 keV) was used to have elemental analysis in UNCD NW film. The scattering angle was kept as 165°.

### B. TRIBOLOGICAL TESTING

Tribological testing on UNCD NW thin film has been carried out using ball-on-disc Micro-tribometer (CSM-Switzerland) working in linear reciprocating mode. The tribological properties of UNCD NW film has been measured while sliding against three different standard ball counterbodies, i.e. $Al_2O_3$, SiC and 100Cr6 steel with diameter 6 mm. The standardized hardness value of these balls is 18 GPa, 26 GPa and 5 GPa, respectively.

The normal constant load of 2 N, sliding speed of 4 cm/sec was used in each experiment. All tests were being run for total sliding distance of 500 m which equals to 62,500 numbers of sliding cycles. The data acquisition rate of 10 Hz and stroke length of 4 mm were used for all experiments. The tests were carried out in ambient (dry and unlubricated conditions).

The wear tracks and ball scars developed after tribology testing were examined using an optical microscope. Three-dimensional wear profiles in tracks were also obtained using 3D optical profilometer (Taylor Hobson).

## III. RESULTS AND DISCUSSIONS

Topography of the diamond film showed homogeneous distribution of diamond particles (Fig. 1a). In the AFM image, agglomeration of diamond particles is not observed. The roughness value of film is ~13 nm. In Raman spectra, a broad peak is deconvoluted into five segments; three of them are related to *TPA* ($v_1$, $v_2$ and $v_3$). These chemical entities are the fingerprint of UNCD structure (Fig. 1b) [4]. G and D band correspond to graphitic and amorphous diamond phase occupying the grain boundary region. Nanowire feature of film is indicated in the FESEM image having high aspect ratio with no agglomeration (Fig 1c). RBS analysis clearly showed that there are two distinct RBS lines related to C and Si peak (Fig. 1d). No other elemental signal was observed which indicates the compositional purity of film. The elastic Modulus and hardness values of film are 165 GPa and 14.5 GPa, respectively.

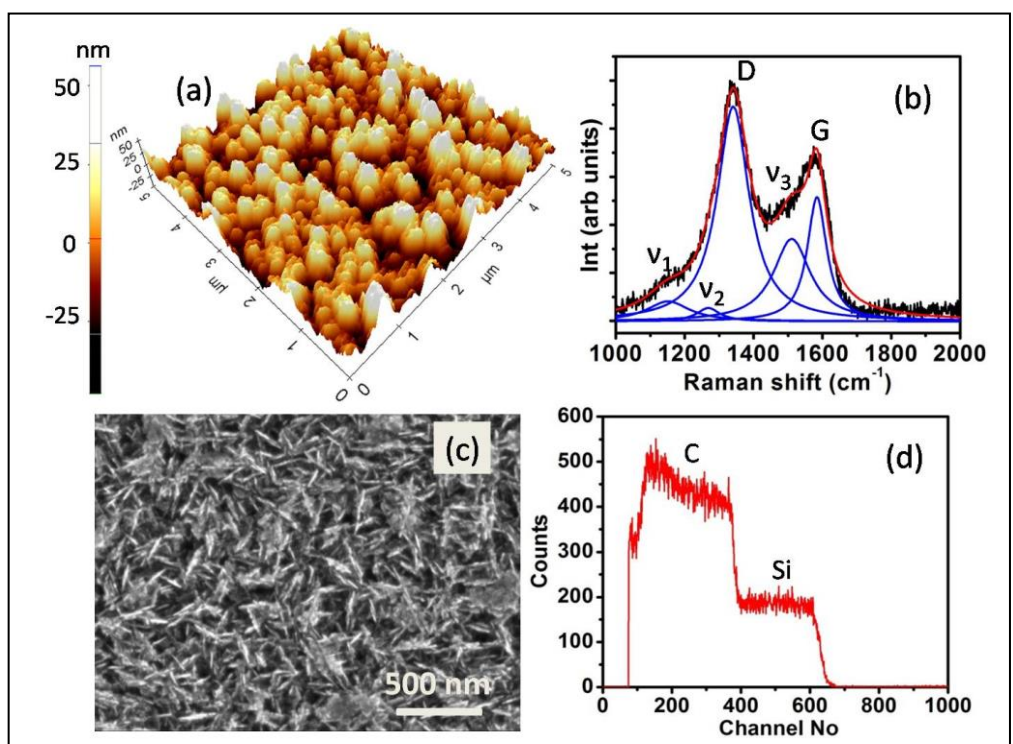

Fig. 1. AFM image of UNCD NW thin film (a), Raman spectra (b), FESEM image (c), and RBS spectra (d).

The friction results of UNCD NW film sliding against various counterbody balls are shown in Fig. 2. Tribotest parameters are mentioned in the inset. In Fig. 3, two-dimensional (2D) wear track and ball scar optical images are shown. In addition, 3D optical image and 2D wear profile of wear track are also shown in Fig. 3.

$Al_2O_3$ ball is chemically inert and thus having weak affinity towards UNCD NW film during sliding process. Hence, the ultralow value of friction coefficient in the range of 0.02-0.06 is maintained throughout the whole sliding process of 500 m (Fig. 2). In the response, ball was also less worn out as shown in Fig. 3a.

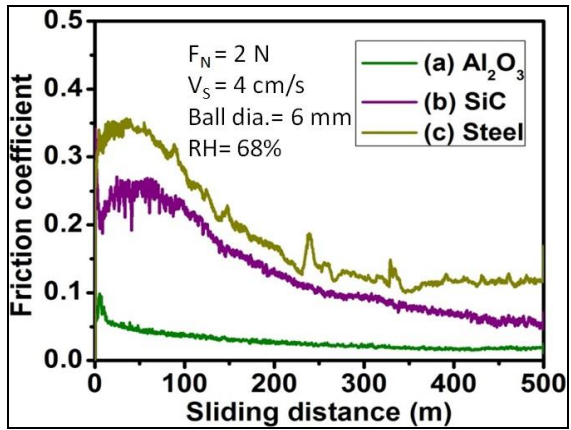

Fig. 2. Friction coefficient of UNCD NW film against $Al_2O_3$, SiC and steel ball.

On the other side, SiC ball is also inert but there is a possibility of C-C strong bond formation across sliding interface with UNCD NW film during the sliding process. Therefore, initially high friction value was observed which gradually decreased to ~0.05 with sliding distance. The average value of friction coefficient in this case is 0.13. In the beginning of the sliding, high friction caused severe wear of ball, scar size ~200 μm (Fig. 3b). Wear debris are clearly observed along edge side of the wear track in the optical image. Conversely, only a few nm of wear depth is seen in the film.

In the case of UNCD NW/steel sliding combination, high friction was observed in the beginning of the sliding cycles which gradually decreased to 0.12. Steel ball is mechanically soft with the hardness value of 5 GPa compared to film having hardness of 14.5 GPa. Therefore, due to soft-hard sliding combination, wear takes place from soft steel ball which develops a transfer layer on film surface as shown in 3D profile of Fig 3c. The wear debris can be observed clearly scattered along edge side in the optical image of wear track (Fig 3c). Contrastingly, in this case, only few nanometer of wear loss was observed from the film.

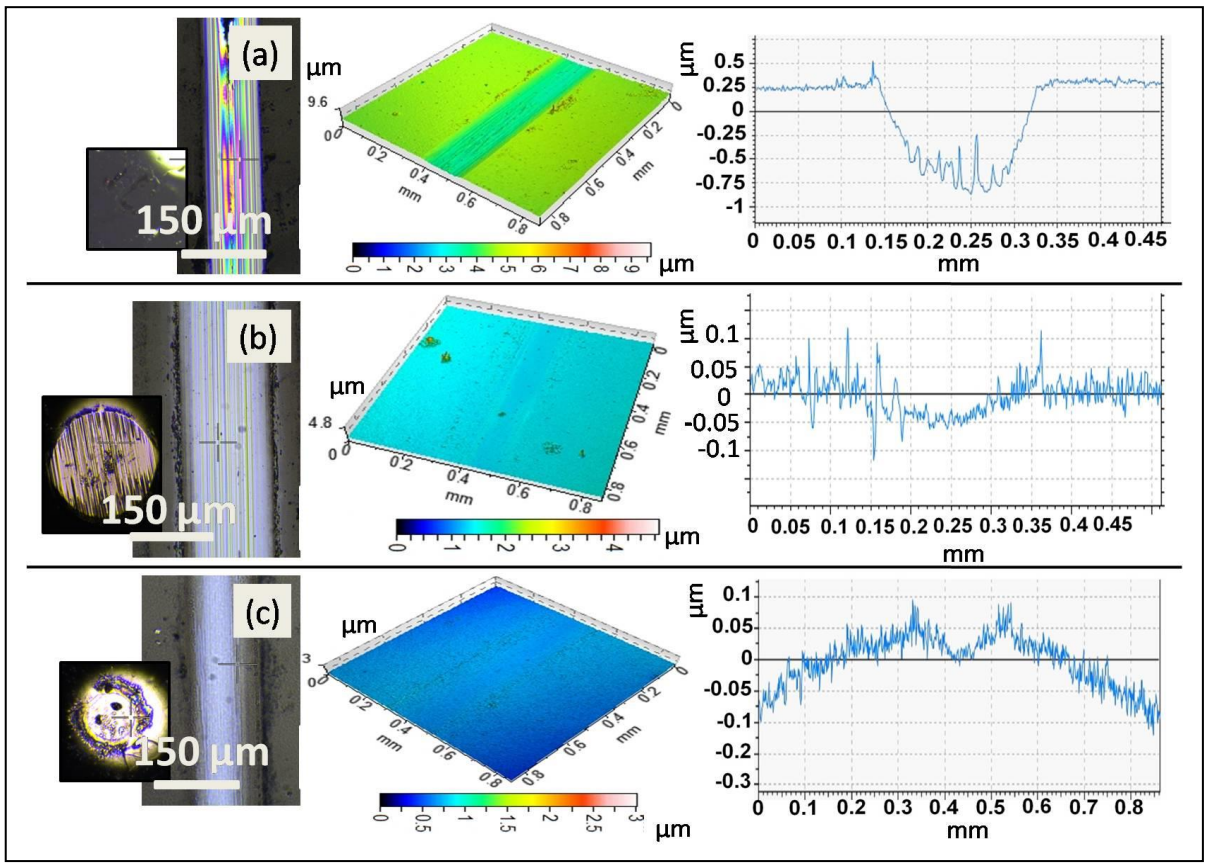

Fig. 3. Wear track and ball scar optical images along with 3D profile of wear track against (a) $Al_2O_3$ (b) SiC and (c) Steel ball, respectively.

From the above results, wear resistance of the film is superior in UNCD NW/SiC and UNCD NW/Steel sliding combination. However, wear loss from the ball was significant due to high frictional energy in the run-in regime. In contrast, UNCD NW/$Al_2O_3$ sliding pair showed negligible wear from the ball counterbody.

This can be possibly due to low friction value in such sliding combination from the beginning itself. But, deep wear track was developed in this case that could be related to mechanical properties of the ball.

## IV. CONCLUSIONS

UNCD NW thin film was deposited on silicon substrate. Tribological studies were carried out against three different balls. In the beginning of sliding process, film friction coefficient against SiC and steel ball showed high value and it was gradually decreased to low value 0.05 and 0.12, respectively. However, ultralow friction value was observed in the sliding combination of UNCD NW/$Al_2O_3$ ball. The wear resistance of film was high in the case of UNCD NW/SiC and UNCD NW/Steel sliding pair. However, wear of the ball was negligible in the case of UNCD NW/$Al_2O_3$ sliding combination.

ACKNOWLEDGMENT

The authors would like to acknowledge Mr. Syamala Rao Polaki and Dr. S. K. Srivastava, IGCAR/Kalpakkam; Dr. D. Dinesh Kumar, Sathyabama University/Chennai and Mr. Pankaj Kr. Das, NIT/Agartala for various experimental help and fruitful discussions.

REFERENCES

[1] Revati Rani, N. Kumar, A.T. Kozakov, K.A. Googlev, K.J. Sankaran, P.K. Das, S. Dash, A.K. Tyagi and I-Nan Lin, "Superlubrication properties of ultrananocrystalline diamond film sliding against a zirconia ball" RSC Adv., vol. 5, pp. 100663-100673, 2015.

[2] N. Kumar, K. Panda, S. Dash, J.P. Reithmaier, B.K. Panigrahi, A.K. Tyagi and B. Raj, "Tribological properties of nanocrystalline diamond films deposited by hot filament chemical vapor deposition" AIP Adv., vol. 2, pp. 032164-032156, 2012.

[3] K. Panda, N. Kumar, K.J. Sankaran, B.K. Panigrahi, S. Dash, H-C. Chen, I-Nan Lin, N-H. Tai and A.K. Tyagi, "Tribological properties of ultrananocrystalline diamond and diamond nanorod films" Surf & Coat Technol., vol. 207, pp. 535-545, 2012.

[4] R. Pfeiffer, H. Kuzmany, P. Knoll, S. Bokova, N. Salk and B. Gunther, "Evidence for trans-polyacetylene in nano-crystalline diamond films" Diamond and Related Mater., vol. 12, pp. 268-271, 2003.

**E-mail of the author(s):** *niranjan@igcar.gov.in,

aggarwalrevati@igcar.gov.in